\documentclass[mathleft
]{an}
\usepackage{graphicx}
\usepackage{times}
\begin{document}

% The following seven commands are intended for editorial usage and should be ignored by
% the author(s).
\Pagespan{1}{6}% Document's page range. 
% If second parameter is left empty, the last page is computed automatically.
\Yearpublication{2012}%
\Yearsubmission{2012}%
\Month{??}%   
\Volume{??}%  
\Issue{??}% 
% \DOI{This.is/not.aDOI}% 

\title{The noise wars in helio- and asteroseismology.}

\author{R.A. Garc\'\i a\inst{1}\fnmsep\thanks{Corresponding author:
  \email{rafael.garcia@cea.fr}\newline}}
\titlerunning{The noise wars}
\authorrunning{R.A. Garc\'\i a}
\institute{Laboratoire AIM, CEA/DSM-CNRS, Universit\'{e} Paris 7 Diderot, IRFU/SAp, Centre de Saclay, 91191, GIf-sur-Yvette, France}

\received{30 August 2012}
\accepted{16 October 2012}
\publonline{??}


\abstract{During this conference, latest results on helioseismology (both local and global) as well as in asteroseismology have been reviewed, the hottest questions discussed and the future prospects of our field fully debated. A conference so rich in the variety of topics adressed is impossible to be deeply reviewed in a paper. Therefore, I present here my particular view of the field as it is today, concentrating on the solar-like stars and global helioseismology. The link I found to do so is the constant battle in which we are all engaged against the sources of noise that difficult our studies. The noise in the data, the noise in the inversions, the precision and accuracy of our inferred models...}

\maketitle

\section{Introduction}

Thanks to the helioseismic ground-based networks that have been operating for 3 decades in the best case (e.g. GONG (Harvey et al. 1996)) and BiSON (Chaplin et al. 1996)); the observational campaigns involving many telescopes all over the world to continuously monitoring stellar pulsations (e.g. the delta scuti network (www.univie.ac.at/tops/dsn/intro.html)), and the long time series provided by the space missions such as, e.g.,  SoHO (Domingo, Fleck, \& Poland 1995), SDO (Schwer et al. 2002),  WIRE (Buzasi et al. 2004), MOST (Matthews 1998), CoRoT (Baglin et al. 2006), and {\it Kepler} (Borucki et al. 2010), we are attending the ``modern era of  Helio- and Asteroseismology'' as stated in the title of the conference. 

The high quality and amount of data sets available and the improvements in the techniques and modeling makes the field to evolve very rapidly and state-of-the-art results today become ``old'' a few months later. It is also important to point out that the situation will be ensured --from the observational side-- for the next years as the {\it Kepler} mission has been extended and CoRoT and SoHO will be very likely extended as well up to the 2016 horizon. However, in this last case, it only concerns the Doppler velocity instrument GOLF (Gabriel et al. 1995; Garc\'\i a et al. 2005) and the photometric instrument VIRGO (Fr\"ohlich et al. 1995) because MDI (Scherrer et al. 1995) has stopped the scientific acquisition in a regular basis in benefit of the newer HMI instrument on board SDO (Scherrer et al. 2012). 

\section{Preparing the data: reducing the noise in the time series.}
The first battle against the noise occurs in the time series. Indeed, instrumentation is not perfect and there is always instrumental noise. On top of that, stars have intrinsic sources of noise --like convection-- making the task of measuring the oscillations more difficult. It has been shown that non-linear time series analyses are powerful tools to reduce this undesirable noise. The idea behind these tools is that all possible states of a system can be represented by a point in a ``phase space'' in which each degree of freedom is an axis of the multidimensional space. The evolution of the states represents the time evolution of the light curve. The phase-space portrait is obtained by plotting light-curve values against values at various time delays. 

\begin{figure*}[!htb]
\includegraphics[width=\linewidth]{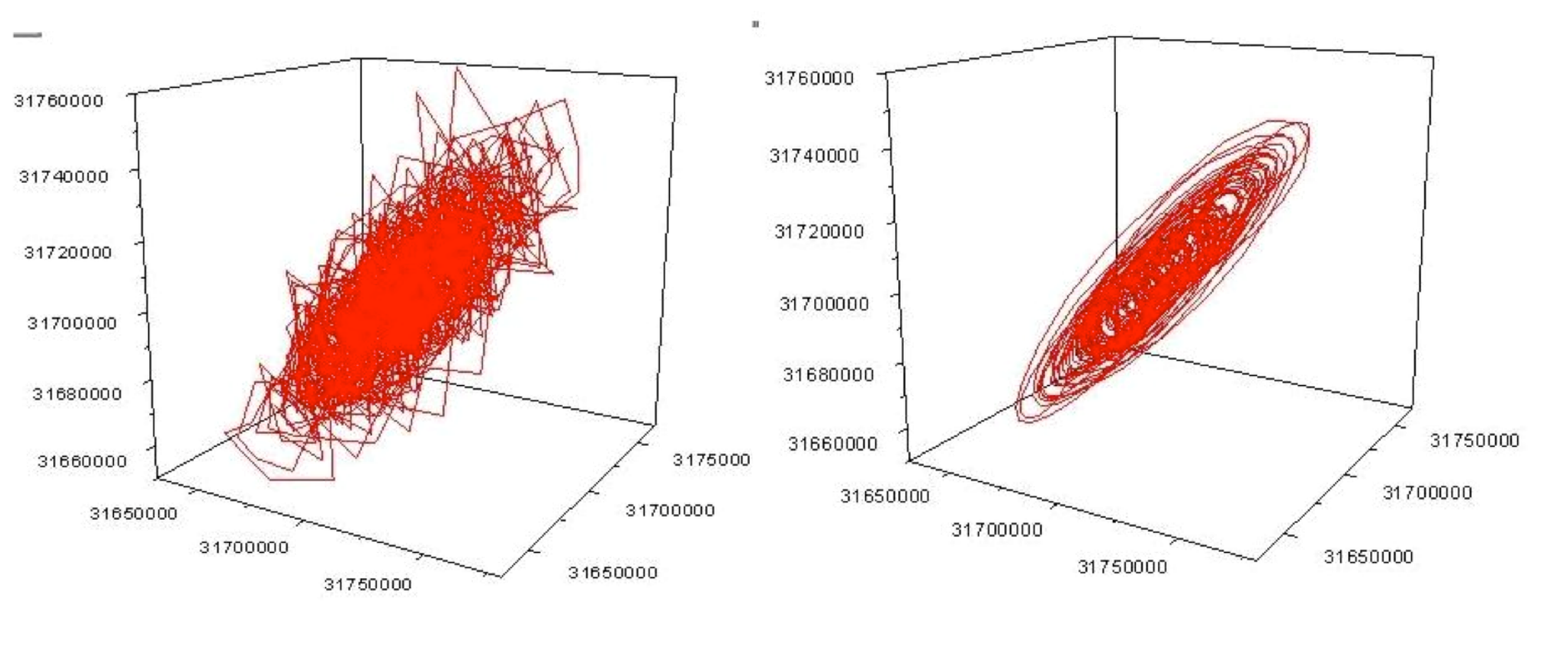}
\caption{Phase-space portraits for KIC~8054179 for the raw data (left) and the data after nonlinear noise reduction (right) for a delay of 1 (for more details see Jevti\'c et al. 2012)}
\label{fig_Nadja}
\end{figure*}

\begin{figure}[!htb]
\includegraphics[width=\linewidth]{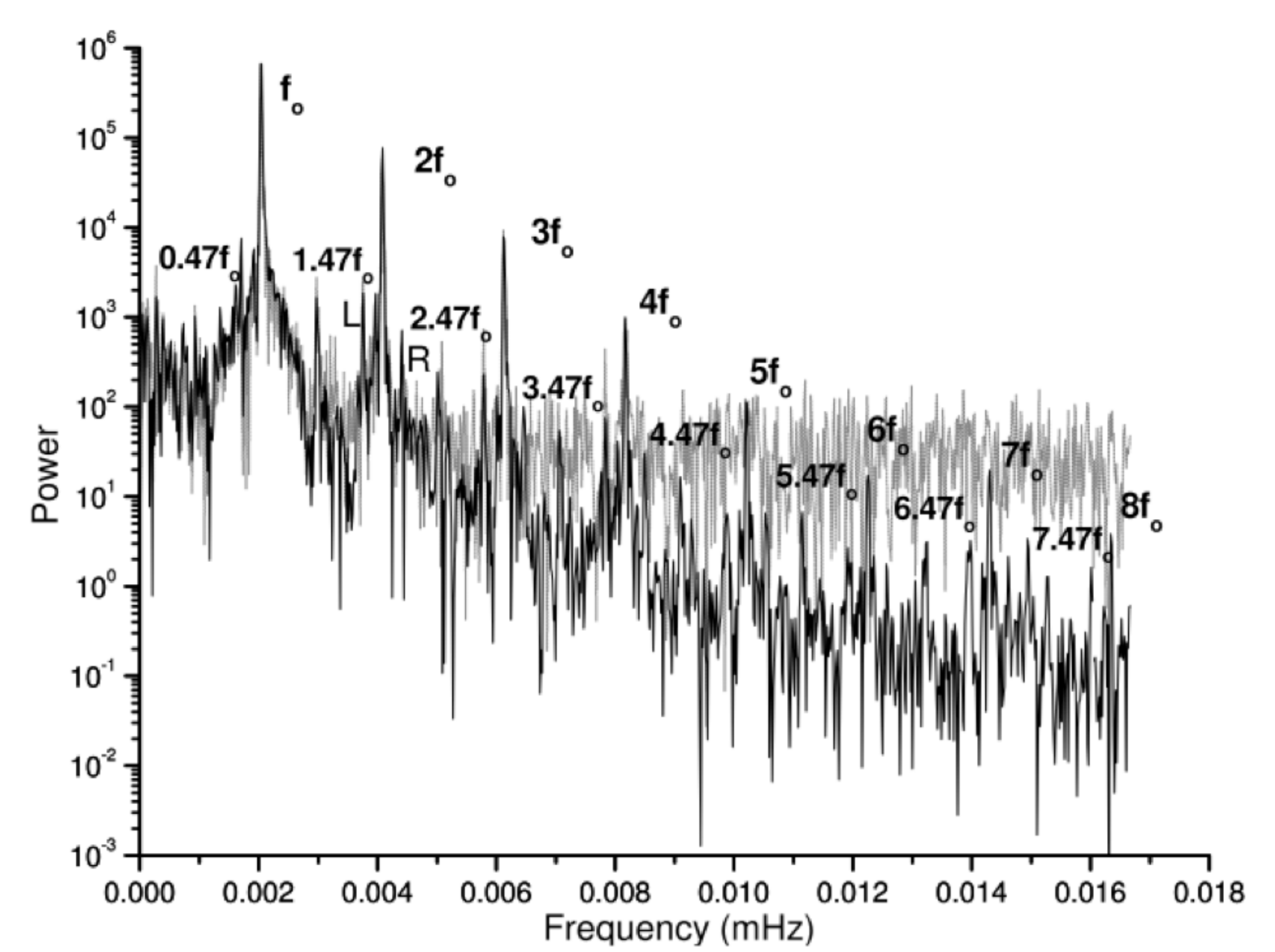}
\caption{Power spectrum of the normal time series (gray) and the one after the non linear reduction (see for more details Jevti\'c et al. 2005).}
\label{fig_Nadja2}
\end{figure}

Figure~\ref{fig_Nadja} represents an example of an efficient noise reduction (right panel) of the more ``chaotic'' raw light curve (left panel). The power spectrum of the filtered time series has a reduction of noise at high frequency that could reach typical values of a few order of magnitude till a factor of $10^4$ in the best cases (see Jevti\'c et al. 2012). Figure~\ref{fig_Nadja2} shows an example of the power spectrum obtained using raw data (gray lines) and the one issued of the local projective nonlinear noise reduction (black lines). At high frequency, when the spectrum is dominated by the noise, the algorithm allows a reduction of about a factor 100 unveiling the existence of more frequencies that were hiden in the raw spectrum.

Another important point we need to take into account when we deal with Fourier analysis is the sampling rate of the data which defines the Nyquist frequency, i.e., the highest frequency of a continuous signal that we can analyze. In the case of {\it Kepler}, most of the data is obtained at a rate of $\sim$~30 minutes (usually called long-cadence data), while at any time 512 targets can be sampled at a shorter cadence of $\sim$~1 minute (see for further details, e.g., Garc\'\i a et al. 2011a). Both light curves can be used for asteroseismic analysis but it is important to be careful when we interpret the spectrum obtained using long-cadenced (LC) data. Indeed Murphy (2012) showed that pulsating frequencies of a delta scuty star with frequencies $\sim 3$ times above the Nyquist frequency could generate a pattern of aliased peaks at frequencies below the Nyquist frequency that could be misinterpreted if only LC data were available.

\section{Extracting the seismic signal from the (noisy) background.}
\begin{figure}
\includegraphics[width=\linewidth]{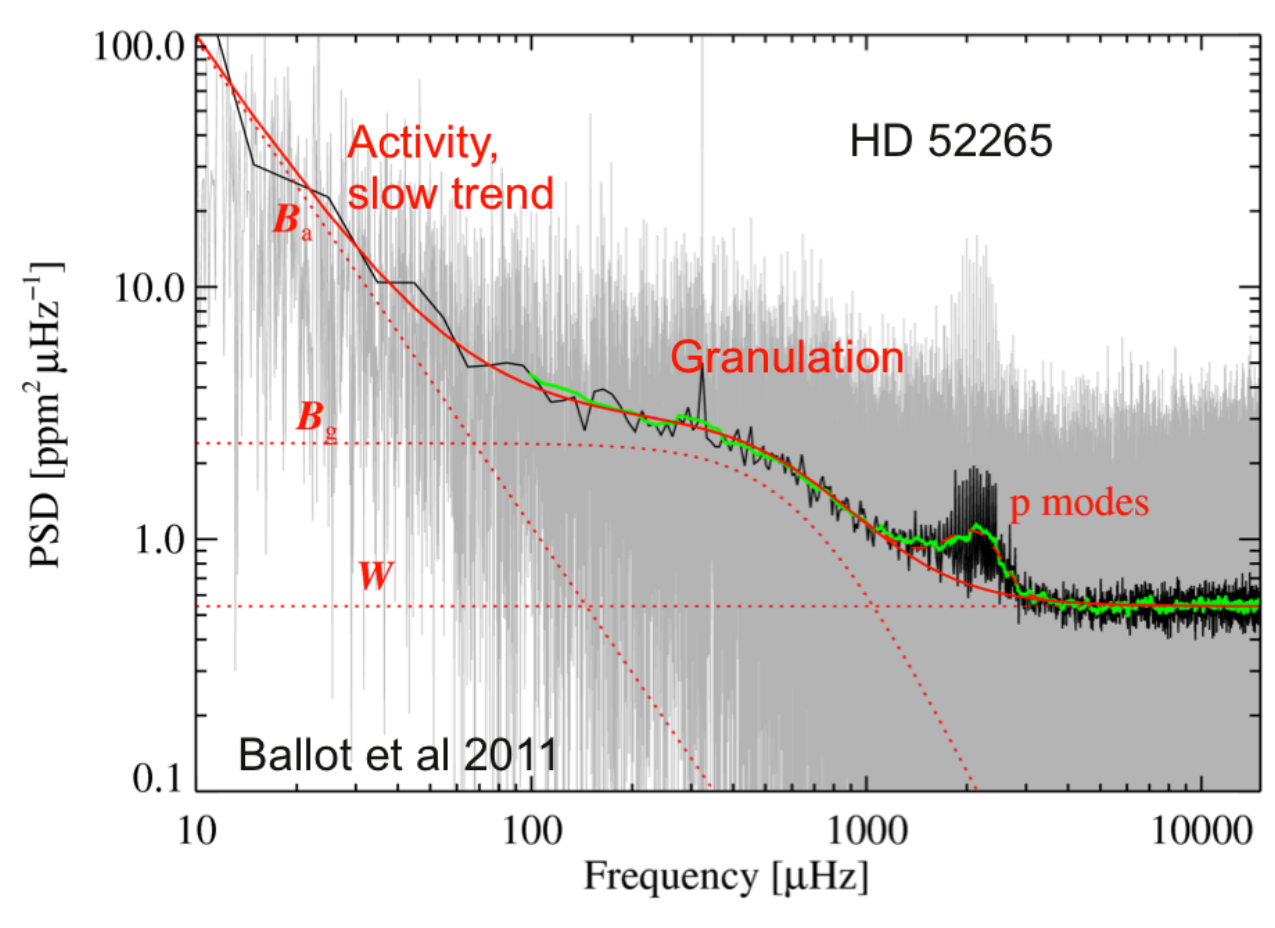}
\caption{PSD of the star HD~52265 observed by CoRoT (adapted from Ballot et al. 2011).}
\label{fig_back}
\end{figure}

In the Fourier domain, the spectrum of many pulsating stars and, in particular, the solar-like stars is dominated by a continuum produced by the turbulent motions induced by the convection. It is usually called the convective background or simply background (e.g. Harvey 1985; Mathur et al. 2011b and references therein). In Figure~\ref{fig_back} the power spectrum density of the star HD~52265 is plotted as a function of the frequency (gray lines). The black and green lines represent smoothed versions of the PSD. The continuous red line represents the background that has been fitted here with 3 components denoted W (the photon noise), the granulation contribution ($B_g$) and an activity component ($B_s$). 

The convection --that generates the background-- also excites the modes of the solar-like oscillating stars playing and important role in the damping, too (Goldreich \& Keeley, 1977). The smaller the lifetime of the modes is, the largest the mode's widths are. When the modes are large, it is more difficult to extract the rotational splitting (see e.g. Fig.~1 of Ballot et al. 2006), complicating the identification of the modes as it is the case in the F stars (e.g. Appourchaux et al. 2008, 2012; Garc\'\i a et al. 2009). Fortunately, using other parameters as the offset $\epsilon$ in the asymptotic relation of the acoustic modes: $\nu_{n,l} \approx \Delta \nu ( n+l/2+\epsilon)-\delta\nu_{0,l}$ (see Vandakurov 1967; Tassoul 1980; Gough 1986), White et al. (2012) were able to correctly identify the  ridges of the modes. Here, $\Delta\nu$ is the large separation between modes of the same harmonic degree $l$ and consecutive radial order n, and $\delta\nu_{0,l}$ is the small separation between modes of different harmonic degree.

As the stars evolve the amplitudes of the oscillations increase and detecting p modes is easier than in young stars (Bedding et al. 2010; Stello et al. 2011; Huber et al. 2011). Indeed, we are able to detect mixed modes in subgiants (e.g Kjeldsen et al. 1995; Deheuvels et al. 2010; Campante et al. 2011; Mathur et al. 2011a) and in red giants (Beck et al. 2011) and thanks to their period spacings, it is  possible to disentangle if a red giant is burning hydrogen in a shell only, or if it has already ignited the central helium too (Bedding et al. 2011; Mosser et al. 2011). Moreover, the high quality of the present datasets available allow us to measure rotational splittings of individual acoustic and mixed modes (Beck et al. 2012; Deheuvels et al. 2012). With this information we have been able to probe the radial differential velocity of some subgiants and red giant stars. Indeed, the dynamics of the deep core of these red giants seems to be more accessible to our investigations than the core of our Sun, for which the general consensus today is that we have only been able to detect the period spacing of the dipole gravity modes but not the individual modes yet (Garc\'\i a et al. 2007, 2008; Appourchaux et al. 2010). However, as time goes by, the signal-to-noise ratio increases and some candidate modes seem to start rising from the background (Garc\'\i a et al. 2011b). A few of them would be enough to infer a proper rotation rate in the core (Mathur et al. 2008). On the theoretical side, Siegel \& Roth (2011) have studied the excitation of stellar oscillations by gravitational waves. If this result is confirmed, it could contribute to increase the detectability of such modes in solar-like stars.

When comparing the inferred radial differential rotation rate with dynamical stellar evolution models (e.g. Pinsonneault et al. 1989; Turck-Chi\`eze et al. 2010) there is a big discrepancy and new physical processes need to be evoked to justify the observational profile. Indeed the modeled rotation is much more steep than the observed ones  (e.g. Marques et al. 2012; Ceillier et al. 2012). In any case, we have now a large sample of stars from CoRoT and {\it Kepler}. Indeed, thousands of red-giant interiors have been proved (e.g. Kallinger et al. 2010; Bedding et al. 2011;  Basu et al. 2011; Mosser et al. 2012a,b), while some stars have already been studied with a greater detail (e.g. Metcalfe et al. 2010a, 2012; di Mauro et al. 2011; Eggenberger et al. 2012; Mathur et al. 2012a).

\section{Local helioseismology}
Compared to the study of oscillation modes trapped in the interior of the Sun (i.e., global helioseismology), local helioseismology's goal is to interpret the full wave field observed at its surface, not just the eigenmode frequencies. Therefore, local helioseismology is able to prove the solar interior in three dimensions, which is important to understand large-scale flows, magnetic structures, and their interactions. Local helioseismology can potentially be used to infer vector flows, thermal and structural inhomogeneities, and even the magnetic field itself (for full reviews on local seismology see:  Duvall Jr 1998;  Kosovichev et al. 2002; Gizon \& Birch 2005; Gizon et al. 2010 and references there in).

Sunspots are manifestations of the presence of magnetic field in the Sun and are also observed in other stars. These structures are one of the typical objects of study of local seismology thanks to its ability to prove the solar interior at different depths as well as the structure and properties of the flows in its neighborhood. Unfortunately, the analysis done with the same technique but  using different instruments or the analysis done with different techniques but the same datasets provide sometimes different results (e.g. Gizon et al. 2009). Therefore, there is a huge amount of time and effort invested by the local helioseismic community to understand the sources of these differences. Realistic models of sunspots are necessary to test local helioseismic techniques and for that we need better observational constraints about their surface structure from classical methods (see Franz, 2012).

\begin{figure*}
\includegraphics[width=\linewidth]{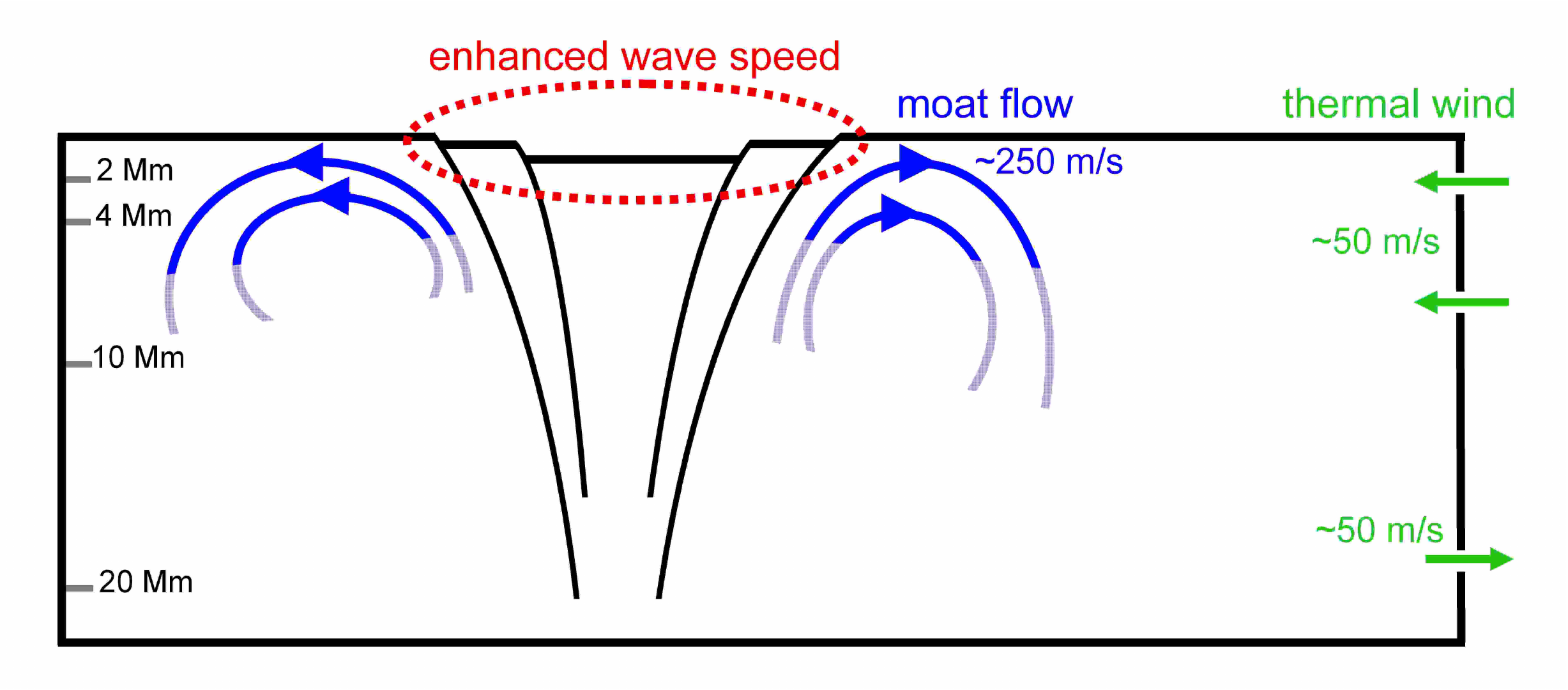}
\caption{Diagram showing the general consensus about the structure of a sunspot (diagram courtesy of L. Gizon).}
\label{fig_Gizon}
\end{figure*}

The general consensus today about sunspots is that wave speed is enhanced in their near-surface layers by the presence of strong magnetic fields (see Figure~\ref{fig_Gizon}). In the immediate vicinity of the sunspot, the 250 m/s outflow in the moat is observed using time-distance helioseismology and ring-diagram analysis. Further away from large active regions (including sunspots or not), a weak inflow of about ~30 m/s is measured near the surface, presumably driven by horizontal pressure gradients. See review article by Gizon, Birch, Spruit (2010) for further explanations.

\section{Classical asteroseismology}
Probably one of the hottest topics today in the classical pulsators astroseismology is the study of hybrid stars (e.g. Belkacem et al. 2009; Grigahc\'ene et al. 2010; Antoci et al. 2011). Indeed, are all the delta scuti and gamma doradus stars hybrids? We cannot answer this question yet but, once again, the high-quality data being collected with the present instrumentation will contribute to shed some light on the problem.

For many classical pulsators, the ``noise'' is not a question of signal-to-noise ratio but to identify the modes among the forest of peaks corresponding to the combination of frequencies, the magnetic and rotation splittings, and any other effect that we could still not know. Frequency and period spacings are powerfull techniques already commonly used in solar-like stars that are now being used in classical pulsators to understand the observed spectrum (e.g.  Garc\'\i a Hern\'andez et al. 2009; Degroote et al. 2012).

In the case of the RR~Lyrae stars, it seems that some stars show very strong cycle-to-cycle changes in its Blazhko modulation, which are caused by both a secondary long-term modulation period and irregular variations (Guggenberger et al. 2012; Moskalik et al. 2012). In all the cases, better models will be required to better understand those stars.

\section{Activity cycles}
The study of magnetic activity cycles in stars observed through seismology is a new way towards the understanding of the physical processes involved in their generation that drives the cycles at different stages of the stellar evolution (e.g. Mathur et al. 2011c; Hekker \& Garc\'\i a 2012). It is commonly accepted that magnetic activity cycles are the consequence of the interaction between the rotation, magnetic fields, and sometimes also on convection. However, the exact way in which they interact is still not perfectly understood (see e.g., Rempel 2008; Charbonneau 2010; Brun \& Strugarek 2012 and references therein). Indeed our lack in the understanding of these processes became clear during the non predicted last extended minimum (e.g. Howe et al. 2009; Salabert et al. 2009).

\begin{figure}[!hb]
\includegraphics[width=\linewidth]{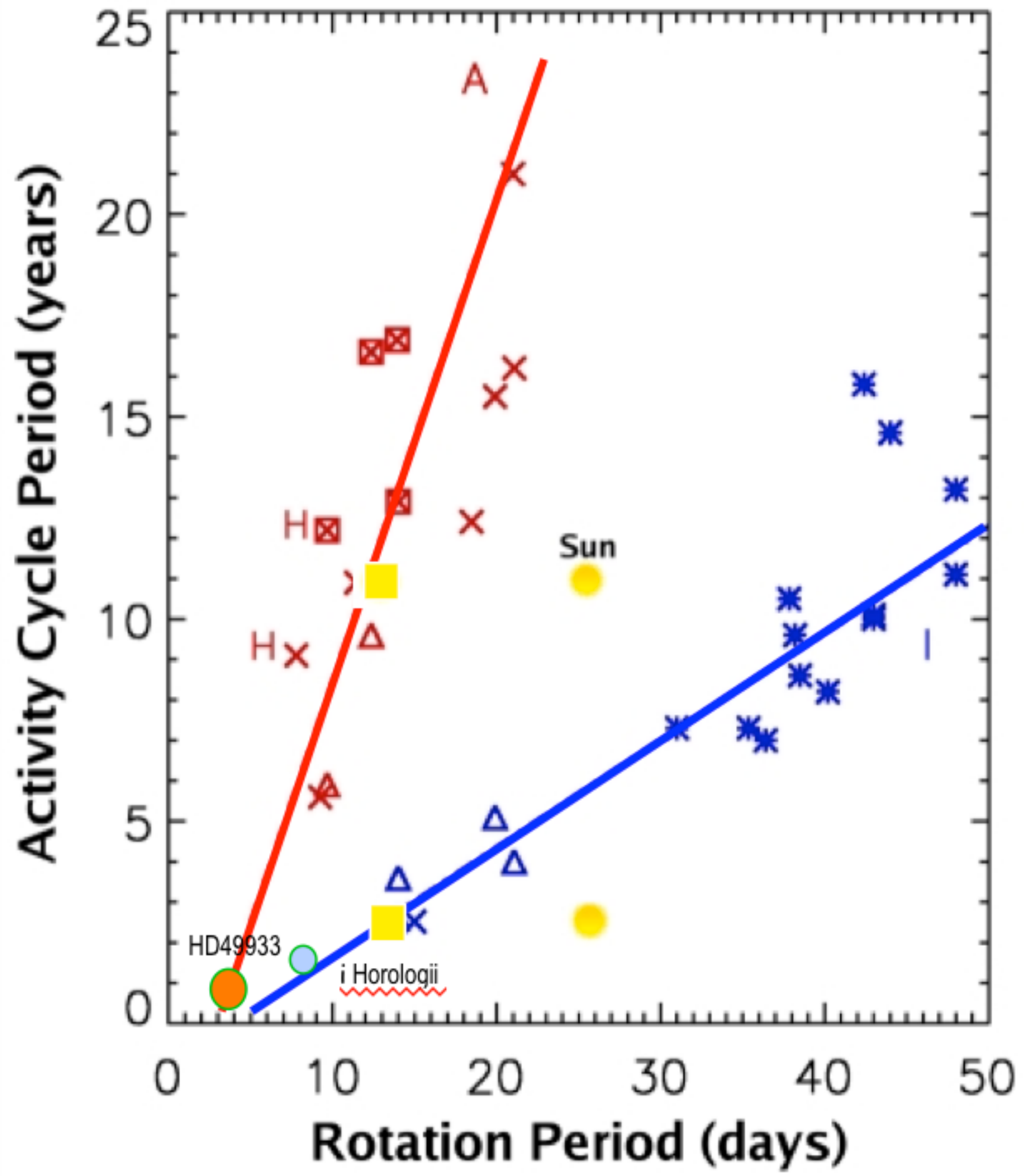}
\caption{Activity cycle period as a function of the rotation period. Two branches are clearly visible, the Active (A) and the inactive (I) one. The yellow circles correspond to the Sun for the 11-year cycle and the possible 2-year one. The yellow squares are the same but considering a rotation period half of the real one. The yellow and light blue circles correspond to HD~49933 and $\iota$ Horologii. (Plot adapted from B\"ohm-Vitense 2007)}
\label{fig_Bohm}
\end{figure}
The importance of asteroseismology is that we can measure at the same time: a) the surface rotation as a consequence of the modulation in the light curve induced by magnetic structures in the photosphere (e.g. Mosser et al. 2009; Mathur et al. 2010; Garc\'\i a et al. 2011c); b) the internal structure of the star, in particular, to extract the depth of the convection zone directly from the data or through out modeling (e.g. Monteiro et al. 2000; Ballot et al. 2004; Mathur et al. 2012); c) the granulation time scale (e.g. Mathur et al. 2011b); d) the seismic magnetic activity proxies (Garc\'\i a et al. 2010).

Stellar activity of hundreds of stars have already been monitored over the last 40 years with the Mount Wilson survey (e.g. Wilson 1978;  Baliunas et al. 1995) or the Lowell Observatory survey (Hall et al. 2007). Many stars seem to have periodic cycles from about a year (Metcalfe et al. 2010b) up to $\sim$ 25 years. Moreover, it has been suggested that there are two different branches of stellar cycles in solar-type stars, one active and one inactive (Saar \& Brandenburg 1999; B\"ohm-Vitense 2007) but one of the problems is the position of the Sun, for which the 11-year solar cycle and the recently discovered 2-year modulation (Fletcher et al. 2010; Broomhall et al. 2011; Simoniello et al. 2012) are shifted compared to the rest of the stars (see yellow circles in Figure~\ref{fig_Bohm}). However, if we plot the solar values at half of the rotation rate (13 days), we obtain a perfect match with the rest of the stars. This is probably a coincidence but when we look to the signature of the sunspots in the solar light curves recorded for example by GOLF/SoHO we do not recover the full solar rotation period but half because we only see half of the Sun. This could explain the discrepancy but when B\"ohm-Vitense applied the methodology to the Sun, the correct rotation period was recovered. The question is then still open and if the signature of many magnetic activity cycles is measured in solar-like stars exhibiting pulsations we will be maybe able to verify this empirical law.

\section{Conclusions}

As said in the title of the conference, we are attending the ``modern era of Helio- and Asteroseismology''. The field has reached its maturity and the results we are obtaining do not only influence the solar and stellar communities  but can also be used by others to feed their research activities. One example of that is the recent asteroseismic studies concerning the stellar population distribution in our galaxy (e.g. Chaplin et al. 2011; Miglio et al. 2012). The field is therefore very active and the data production is ensured by the ground-based facilities as well as by the space instrumentations currently deployed. Finally modeling is being refined and open questions a few years ago such as what is the internal rotation profile of distant stars or how to disentangle between the nature of  red giant stars can now being addressed. We are assisting to the revolution of our field.

\acknowledgements
I thank the organizers of the HelAs V meeting in \"Obergurgl, Austria and the European Science Foundation (ESF)  for the invitation. I also thank L. Gizon and N. Jevti\'c for their useful comments and discussions.

\end{document}